\documentclass[a4paper,fleqn,usenatbib,useAMS]{mnras}

\usepackage{graphicx}	
\usepackage{amsmath}	
\usepackage{amssymb}	
\usepackage{multicol}        
\usepackage{bm}		
\usepackage{pdflscape}	
\usepackage{subfigure} 

\newcommand{\Mjup}{$M_{\rm Jup}$}
\newcommand{\ms}{\mbox{m\,s$^{-1}$}}

\usepackage[T1]{fontenc}
\usepackage{ae,aecompl}

\usepackage{newtxtext,newtxmath}

\title[Stability analysis of three exoplanet systems]{Stability analysis of three exoplanet systems}

\author[J.~P. Marshall et al.]{J.~P. Marshall$^{1,2,}$\thanks{E-mail: jmarshall@asiaa.sinica.edu.au}, J. Horner$^{2}$, R. A. Wittenmyer$^{2}$,  J. T. Clark$^{2}$, M.~W. Mengel$^{2}$\\
$^{1}$Academia Sinica, Institute of Astronomy and Astrophysics, 11F Astronomy-Mathematics Building, NTU/AS campus,\\ No. 1, Section 4, Roosevelt Rd., Taipei 10617, Taiwan\\
$^{2}$University of Southern Queensland, Centre for Astrophysics, Toowoomba, QLD 4350, Australia}

\date{Accepted XXX. Received YYY; in original form ZZZ}

\pubyear{2019}

\begin{document}
\label{firstpage}
\pagerange{\pageref{firstpage}--\pageref{lastpage}}
\maketitle

\begin{abstract}
The orbital solutions of published multi-planet systems are not necessarily dynamically stable on timescales comparable to the lifetime of the system as a whole. For this reason, dynamical tests of the architectures of proposed exoplanetary systems are a critical tool to probe the stability and feasibility of the candidate planetary systems, with the potential to point the way towards refined orbital parameters of those planets. Such studies can even help in the identification of additional companions in such systems.

Here we examine the dynamical stability of three planetary systems, orbiting HD~67087, HD~110014, and HD~133131A. We use the published radial velocity measurements of the target stars to determine the best-fit orbital solutions for these planetary systems using the \textit{Systemic} console. We then employ the $n$-body integrator {\sc Mercury} to test the stability of a range of orbital solutions lying within 3-$\sigma$ of the nominal best-fit for a duration of 100 Myr. From the results of the $n$-body integrations, we infer the best-fit orbital parameters using the Bayesian package {\sc Astroemperor}. We find that both HD~110014 and HD~133131A have long-term stable architectures that lie within the 1-$\sigma$ uncertainties of the nominal best-fit to their previously determined orbital solutions. However, the HD~67087 system exhibits a strong tendency toward instability on short timescales. We compare these results to the predictions made from consideration of the angular momentum deficit criterion, and find that its predictions are consistent with our findings.
\end{abstract}




\begingroup
\let\clearpage\relax
\endgroup
\newpage

\section{Introduction}
\label{sec:Int}

Exoplanets -- planets orbiting stars other than the Sun -- are most often identified through indirect means. We observe a star with periodic behaviour that would otherwise be unexpected and conclude that the best explanation is the presence of one (or more) planets. We direct the interested reader to \cite{2018Perryman} for a summary of the various exoplanet detection techniques. By piecing together the observations of the unexpected behaviour it is possible to constrain, to some degree, the orbit and physical nature of the planets in question. Such inference is, however, not perfect -- particularly when the planets in question have been detected through observations of the ``wobble'' of their host star, as is the case for planets found using the radial velocity technique \citep[e.g. ][]{51peg,47UMa,HarpsM,114613,3167c}, or candidate planets claimed on the basis of binary star eclipse timing variability \citep[e.g. ][]{hwvir,nnser,silly1,uzfor}.

The accurate determination of the (minimum) masses and orbits of newly discovered exoplanets provides the key data by which we can understand the variety of outcomes of the planet formation process. As such, it behooves us to ensure that exoplanet catalogues contain information which is as accurate and realistic as possible. Such accurate solutions do not just enable us to properly ascertain the distribution of planets at the current epoch -- they also provide an important window into the history of the planetary systems we discover \citep[e.g. ][]{2014Ford,2015PuWu,2017Fulton,2019Wu}, and allow us to predict and plan follow up observations through population synthesis models \citep[e.g. ][]{2013Hasegawa,2018Mordasini,2020Dulz}. For example, the migration and mutual gravitational interaction of planets have been identified as being of critical importance to both the observed architectures and predicted long-term stability of the menagerie of known multi-planet systems heretofore identified through radial velocity and transit surveys \citep[e.g.][]{prototrap,2012cWittenmyer,chain,trappist,2017Mustill,2017Hamers,2019Childs}. 

However, the accuracy of orbital parameters of the planetary companions presented in discovery works is frequently limited by the time period covered by the observations that led to the discovery, which are often enough to claim detection and little more \citep{JupiterAnalogues,CoolJupiters}. Long term follow-up of known planet host systems is therefore desirable to refine the orbital parameters for known companions, to infer the presence of additional companions at lower masses and/or larger semi-major axes \citep[e.g.][]{2017BeckerAdams,EightSingle,RVDItech,181433,2019Denham,Rick19}, and to disentangle the complex signals produced by planets on resonant \citep[e.g.][]{Ang10,2012cWittenmyer,2014aWittenmyer,2016Wittenmyer} or eccentric orbits \citep[e.g.][]{2013Wittenmyer,2017cWittenmyer}. Equally, due to the relative paucity of data on which planet discoveries are often based, it is possible for those initial solutions to change markedly as more data are acquired. The ultimate extension of this is that, on occasion, the process by which planetary solutions are fit to observational data can yield false solutions -- essentially finding local minima in the phase space of all possible orbital solutions that represent a good theoretical fit to the data whilst being unphysical. It is therefore important to check the dynamical feasibility of multi-planet solutions that appear to present a good fit to observational data -- particularly in those cases where such solutions invoke planets on orbits that offer the potential for close encounters between the candidate planets.

Following this logic, we have in the past tested the stability of multi-planet systems in a variety of environments including around main sequence \citep{2010Marshall,2012bWittenmyer,2015Wittenmyer,2017bWittenmyer}, evolved \citep{2017aWittenmyer,2017cWittenmyer,2019Marshall}, and post-main sequence stars \citep{2011Horner,2012Horner,QSVir,2012aWittenmyer,2013Mustill}. In some cases, our results confirmed that the proposed systems were dynamically feasible as presented in the discovery work, whilst in others, our analysis demonstrated that alternative explanations must be sought for the observed behaviour of the claimed ``planet-host'' star \citep[e.g.][]{2011Horner,QSVir}. To ensure that our own work remains robust, we have incorporated such analysis as a standard part of our own exoplanet discovery papers. We test all published multi-planet solutions for dynamical stability before placing too great a confidence in a particular outcome. As an extension to this approach, we presented a revised Bayesian method to the previously adopted frequentist stability analysis in \cite{2019Marshall}, and demonstrated the consistency between these approaches. 

Rather than using direct dynamical simulations, the stability of a planetary system can also be inferred from a criterion derived from the planetary masses, semi-major axes, and conservation of the angular momentum deficit \citep[AMD,][]{2000Laskar,2017Laskar}. AMD can be interpreted as measuring the degree of excitation of planetary orbits, with less excited orbits implying greater stability. The definition of AMD stability has been revised to account for the effect of mean motion resonances and close encounters on orbital stability \citep{2017Petit,2018Petit}. Of the systems examined in this work, HD~110014 has been identified as being weakly stable, whilst HD~67087 and HD~133131A are both considered unstable according to AMD \citep[see Figs. 6 and 7,][]{2017Laskar}. In our previous dynamical studies, we find good agreement between the stability inferred from AMD and our dynamical simulations with 13 systems in common between them, of which nine were classified unstable and four stable, one marginally so \citep{2012Horner,2012Robertson,2012aWittenmyer,2012bWittenmyer,2012cWittenmyer,2014aWittenmyer,2014bWittenmyer,2015Wittenmyer,2016Wittenmyer,2016Endl}. In this paper we examine the dynamical stability of the three multi-planet systems, HD~67087, HD~110014, and HD~133131A, as a critical examination of their stability and a further test of the reliability of AMD for the identification of instability in exoplanet systems. 

HD~67087, observed as part of the Japanese Okayama Planet Search programme \citep{2005Sato}, was discovered to host a pair of exoplanets by \cite{2015Harakawa}. The candidate planets are super-Jupiters, with $m~\sin i$ of 3.1\,\Mjup\ and 4.9\,\Mjup, respectively.  They move on orbits with (${a,e}$) of ({1.08, 0.17}) and ({3.86, 0.76}), respectively, which would place the outer planet amongst the most eccentric Jovian planets identified thus far. The authors noted that the orbit and mass of the outer planet are poorly constrained.   

HD\,110014 was found to host a planet by \citet{2009deMedeiros}; the second companion was identified through re-analysis of archival spectra taken by the FEROS instrument \citep{1998KauferPasquini} looking to derive an updated orbit for planet b \citep{2015Soto}.  The two candidate planets have super-Jupiter masses, and \citet{2015Soto} cautioned that the proposed second planet was worryingly close in period to the typical rotation period of K giant stars. However, their analysis of the stellar photometry was inconclusive in identifying its activity as the root cause for the secondary signal. 

HD~133131A's planetary companions were reported in \citet{2016Teske}, based on precise radial velocities primarily from the Magellan Planet Finder Spectrograph \citep{2006Crane,2008Crane,2010Crane}. Their data supported the presence of two planets, where the outer planet is poorly constrained due to its long period. \citet{2016Teske} ran a single dynamical stability simulation on the adopted solution and found it to remain stable for the full $10^5$ yr duration. The authors presented both a low- and high- eccentricity solution, reasoning that in a formal sense, the two solutions were essentially indistinguishable. They favoured the low-eccentricity model ($e_2=0.2$) for dynamical stability reasons. There is precedent in the literature for this choice, since it does happen that the formal best fit can be dynamically unfeasible whilst a slightly worse fit pushes the system into a region of stability \citep[e.g.][]{2013Mustill, 2014Trifonov, 2017bWittenmyer}.  

The remainder of the paper is laid out as follows. We present a brief summary of the radial velocity observations and other data (e.g. stellar parameters) used for our reanalysis in Sect. \ref{sec:Obs} along with an explanation of our modelling approach. The results of the reanalyses for each target are shown in Sect. \ref{sec:Res}. A brief discussion of our findings in comparison to previous work on these systems is presented in Sect. \ref{sec:Dis}. Finally, we present our conclusions in Sect. \ref{sec:Con}.

\section{Observations and methods}
\label{sec:Obs}

\subsection{Radial Velocity Data}

We compiled radial velocity values from the literature for the three systems examined in this work; the origin of these data are summarised in Table \ref{tab:rvs}. 

\begin{table}
\centering
\caption{Table of references for the radial velocities data used in this work. \label{tab:rvs}}
\begin{tabular}{ll}
    \hline
    Target & References \\
    \hline\hline
    HD~67087   & \cite{2015Harakawa} \\
    HD~110114  & \cite{2009deMedeiros} \\
    HD~133131A & \cite{2016Teske} \\
    \hline
\end{tabular}
\end{table}

\begin{table*}
\centering
\caption{Planetary orbital parameters based on \textit{Systemic} fits to radial velocity data. Semi-major axes were calculated using measured orbital periods and stellar masses taken from the NASA Exoplanet Archive. \label{tab:systemic}}
\begin{tabular}{lcccccc}
    \hline
    & HD\,67087 b & HD\,67087 c & HD\,110014 b & HD\,110014 c & HD\,133131A b & HD\,133131A c \\
    \hline \hline
    Amplitude [\ms] &  74.0~$\pm$~3.0 & 54.0~$\pm$~4.0 & 36.949$\pm$0.750 & 5.956$\pm$1.617 & 135.315$\pm$3.640 & 64.660$\pm$3.966 \\
    Period [days] & 352.3$\pm$1.7 & 2380$^{+167}_{-141}$ & 877.5$\pm$5.2 & 130.125$\pm$0.096 & 648.$\pm$3 & 5342$^{+7783}_{-2009}$ \\
    Mean anomaly [deg] & 35$^{+20}_{-16}$ & 94$^{+28}_{-35}$ & 155$\pm$4 & 231$\pm$3 & 265$\pm$11 & 188$^{+104}_{-123}$ \\
    Longitude [deg] & 281$^{+18}_{-15}$ & 256 (fixed) & 41$\pm$3 & 302$\pm$4 & 16.6$^{+4.7}_{-4.5}$ & 110$^{+25}_{-43}$ \\
    Eccentricity & 0.18$^{+0.07}_{-0.06}$ & 0.51 (fixed) & 0.259$\pm$0.017 & 0.410$\pm$0.022 & 0.340$\pm$0.032 & 0.63$^{+0.25}_{-0.20}$ \\
    $M$ sin $i$ [$M_{\rm Jup}$] & 3.10$^{+0.15}_{-0.14}$ & 3.73$^{+0.47}_{-0.45}$ & 10.61$\pm$0.25 & 3.228$\pm$0.098 & 1.418$\pm$0.036 & 0.52$^{+0.45}_{-0.17}$ \\
    Semi-major Axis [au] & 1.08 & 3.87 & 2.32 & 0.65 & 1.44 & 5.88 \\
    \hline
\end{tabular}
\end{table*}


\begin{table*}
\centering
\caption{Results from \textsc{Astroemperor} exploration of parameter space around \textit{Systemic} nominal best fit values for planetary companions to HD~133131A and HD~110014. \label{tab:mcmc}}
\begin{tabular}{lcccc}
    \hline
    & HD\,133131A b & HD\,133131A c & HD\,110014 b & HD\,110014 c \\
    \hline \hline
    Amplitude [\ms] & 36.949$\pm$0.750 & 5.956$\pm$1.617 & 135.315$\pm$3.640 & 64.660$\pm$3.966 \\
    Period [days] & 647.816$\pm$1.575 & 3205.648$\pm$948.063 & 865.206$\pm$6.170 & 132.431$\pm$0.279 \\
    Phase [deg] & 261.620$\pm$4.850 & 31.734$\pm$98.433 & 43.753$\pm$72.179 & 341.373$\pm$64.346 \\
    Longitude [deg] & 18.550$\pm$2.165 & 113.777$\pm$81.302 & 146.633$\pm$17.229 & 236.903$\pm$16.452 \\
    Eccentricity & 0.341$\pm$0.021 & 0.263$\pm$0.145 & 0.011$\pm$0.015 & 0.294$\pm$0.076 \\
    M sin $i$ [$M_{\rm Jup}$]& 1.428$\pm$0.099 & 0.388$\pm$0.124 & 10.622$\pm$0.757 & 2.581$\pm$0.247 \\
    Semimajor Axis [au] & 1.435$\pm$0.046 & 4.153$\pm$0.800 & 2.350$\pm$0.075 & 0.668$\pm$0.023 \\
    \hline
    Jitter [\ms] & 3.557$\pm$1.254 & 0.466$\pm$0.419 & 6.060$\pm$1.856 & 13.350$\pm$1.492 \\
    Offset [\ms] & -9.333$\pm$4.787 & 12.321$\pm$7.543 & 52.575$\pm$4.737 & 72.198$\pm$4.541 \\
    MA coefficient & 0.714$\pm$0.531 & 0.466$\pm$0.419 & 0.697$\pm$0.214 & 13.350$\pm$1.492 \\
    MA Timescale [days] & 4.158$\pm$2.815 & 12.321$\pm$7.543 & 9.793$\pm$2.488 & 72.198$\pm$4.541 \\
    Acceleration [\ms/yr] & -1.435 & & -21.620 & \\
    \hline
\end{tabular}
\end{table*}

\subsection{Modelling}

To test the dynamical stability of these proposed planetary systems, we follow the updated dynamical methodology outlined in our previous work \citep[][]{2017bWittenmyer,2019Marshall}.

In brief, we perform a fit to the published velocity data using the \textit{Systemic Console} \citep{2009Meschiari}. We then use the MCMC tool within \textit{Systemic} to explore the parameter space about the best fit. The MCMC chain runs for $10^7$ steps, discarding the first 10,000, and we then draw the trial solutions for our dynamical stability simulations from these posteriors. Using this data, we populate three ``annuli'' in $\chi^2$ space corresponding to the ranges $0-1\sigma$, $1-2\sigma$, and $2-3\sigma$ from the best fit.  Each annulus contains 42,025 unique realisations drawn from the MCMC chain. The innermost annulus was drawn from the lowest 68.3 per cent of all $\chi^2$ values, the middle annulus contained the next best 27.2 per cent of values, and the outer annulus contained the worst 4.5 per cent of solutions (i.e. those falling $2-3\sigma$ away from the best fit). The result is a set of ``clones'' which fall within $3\sigma$ of the best-fit solution, thus representing a reasonable region of parameter space within which we explore the dynamical stability of the proposed planetary system, using the constraints afforded by the existing observational data.

We then proceed to perform lengthy dynamical simulations of each of the 126,075 solutions generated by this method. We used the Hybrid integrator within the $n$-body dynamics package \textit{Mercury} \citep{1999Chambers} to integrate the solutions forwards in time for a period of 100 Myr. The simulations are brought to a premature end if either of the planets being simulated is ejected from the system, is flung in to the central star, or if the two planets collide with one another. When such events occur, the time at which the collision or ejection occurred is recorded, giving us the lifetime for that particular run. As such, our suite of simulations yield 126,075 tests of the candidate planetary system, allowing us to study how its stability varies as a function of the particular details of the solution chosen to explain the observational data.

We determine the best-fit parameters and uncertainties for each system using the code Exoplanet Mcmc Parallel tEmpering Radial velOcity fitteR\footnote{\href{https://github.com/ReddTea/astroEMPEROR}{https://github.com/ReddTea/astroEMPEROR}} ({\sc astroEMPEROR}), which uses thermodynamic methods combined with MCMC. Our approach has previously been established and described in \cite{2019Marshall} and \cite{EightSingle}. We summarise the input values and constraints used in the fitting presented in this work for the sake of reproducibility. Given that our goal was to test the feasibility of the exoplanetary systems as presented in the literature, we restricted {\sc astroEMPEROR} to consider zero, one, or two planetary signals in the radial velocity data; dynamical configurations with additional planetary companions in orbits that could mimic a single planetary companion, e.g. two resonant planets looking like a single eccentric planet (for a total of three planetary companions), were not considered in this analysis. The planetary fitting parameters were the orbital period ($P$), line-of-sight mass ($M\sin i$), orbital eccentricity ($e$), longitude of periastron ($\omega$), and mean anomaly ($M$). We also include an additional jitter term when fitting the data. We initialised the locations of the walkers in the MCMC fitting at their best-fit values from the \textit{Systemic} console fit, plus a small random scatter. The priors on each parameter were flat and unbounded i.e. with uniform probability between $\pm\infty$, except for the orbital eccentricities which had folded Gaussian priors, and the jitter term, which was a Jeffries function (but still unbound between $\pm\infty$). The parameter space was surveyed by 150 walkers at five temperatures over 15\,000 steps, with the first 5\,000 steps being discarded as the burn-in phase.

\section{Results}
\label{sec:Res}

\subsection{HD~67087}

\begin{figure*}
    \includegraphics[width=0.45\textwidth]{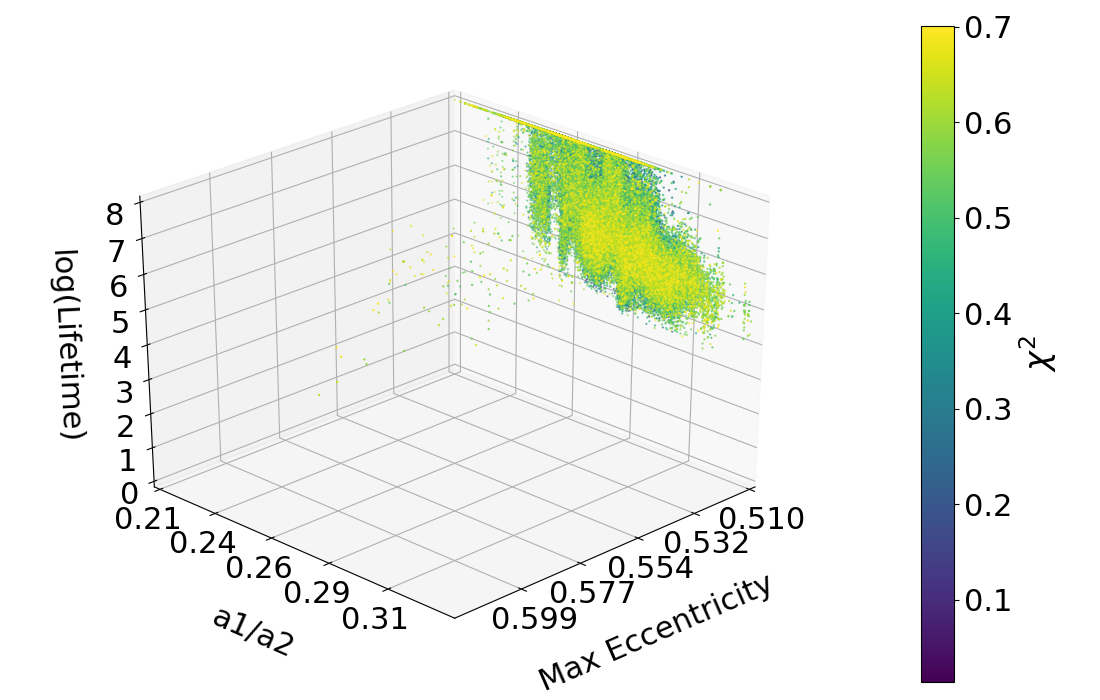}
	\includegraphics[width=0.45\textwidth]{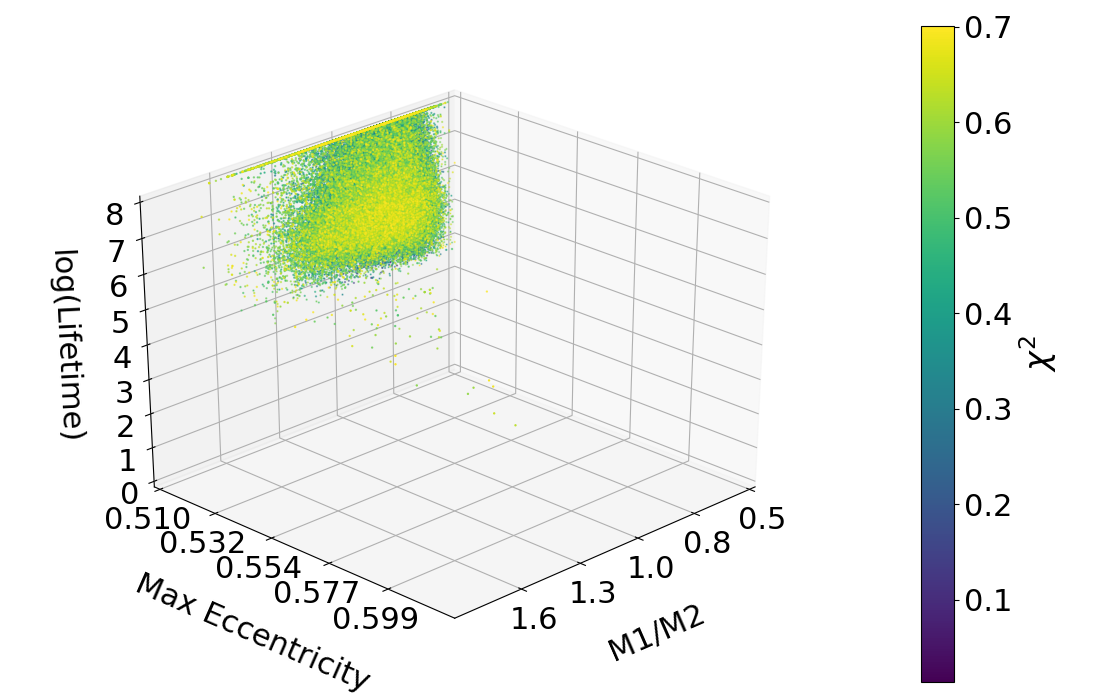}
    \caption{Visualisation of the dynamical stability of the HD~67087 planetary system. On the left we show the log(lifetime) as a function of the largest initial eccentricity fit to HD~67087b and c and the ratio of their orbital semi-major axes, whilst on the right we show the log(lifetime) as a function of of the largest initial eccentricity fit and the mass ratio between HD~67087b and c. The colour bar shows the goodness of fit ($\chi^{2}$) of each solution tested. We find no stable solutions that last the full 100 Myr duration of the dynamical simulations close to the nominal best-fit orbital solution for the planets, with the only stable solutions lying at the extreme edges of the parameter space toward low eccentricities, large separations and low mass ratios. \label{fig:HD67087_stability}}
\end{figure*}

The HD~67087 system is catastrophically unstable, as illustrated by the results of our stability analysis in Fig. \ref{fig:HD67087_stability}. In this plot it is clear that the most stable solutions cluster toward the largest ratios of semi-major axes, and the smallest eccentricities. Even in this limit, the longest-lived solutions that plausibly represent the observations are still only stable for 10$^{6}$ yrs, out of a total integration time of 10$^{8}$ yrs. This leads us to the interpretation that the HD~67087 system, as inferred from the available radial velocity data, is dynamically infeasible. Given this high degree of instability, we do not attempt to determine a global best-fit solution for the system parameters.

\subsection{HD~110014}

\begin{figure*}
    \includegraphics[width=0.45\textwidth]{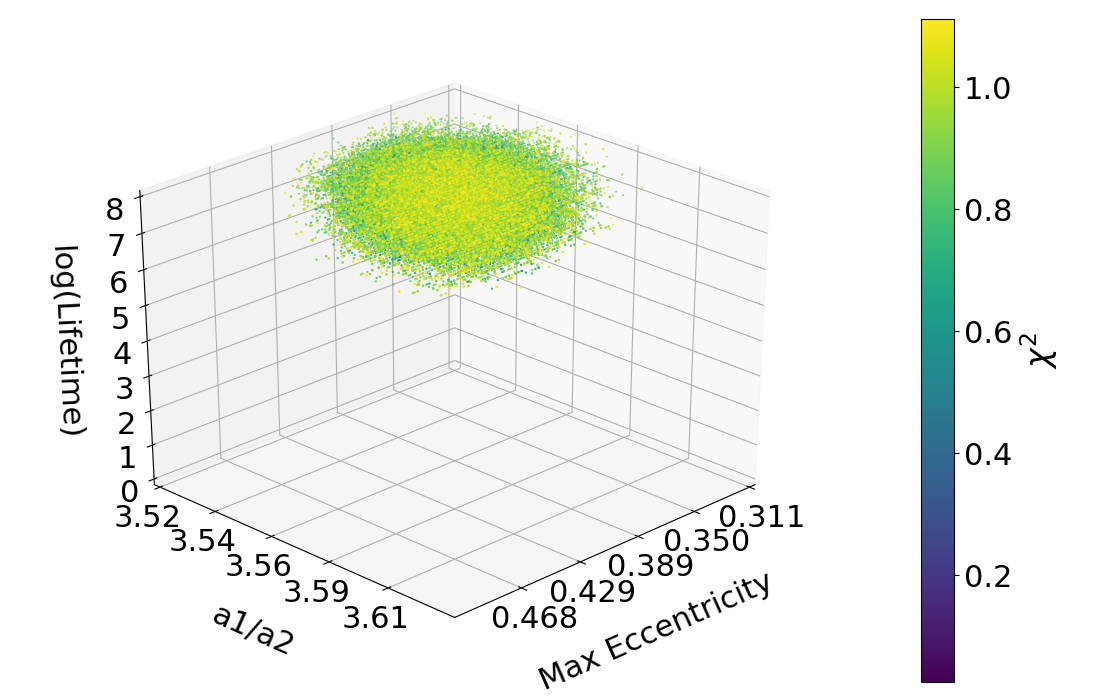}
	\includegraphics[width=0.45\textwidth]{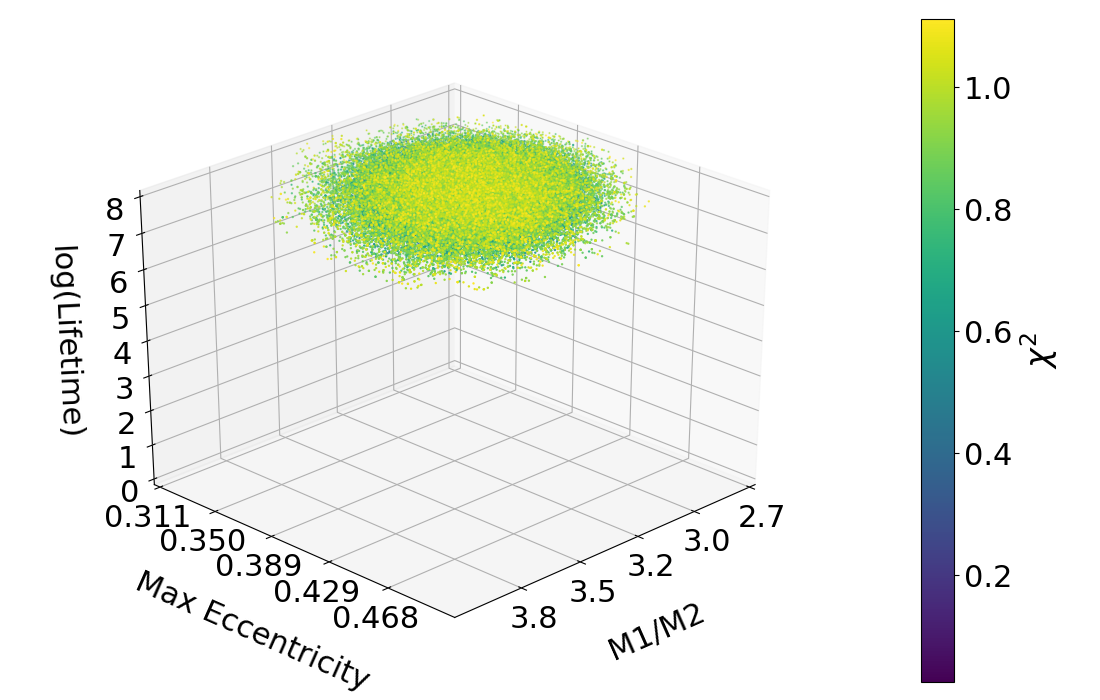}
    \caption{Visualisation of the dynamical stability of the HD~110014 planetary system. On the left we show the log(lifetime) as a function of the largest initial eccentricity fit to HD~110014b and c and the ratio of their orbital semi-major axes, whilst on the right we show the log(lifetime) as a function of of the largest initial eccentricity fit and the mass ratio between HD~110014b and c. The colour bar shows the goodness of fit ($\chi^{2}$) of each solution tested. We find stable solutions that last the full 100 Myr duration of the dynamical simulations close to the nominal best-fit orbital solution for the planets. \label{fig:HD110014_stability}}
\end{figure*}

\begin{figure*}[h]
	\includegraphics[width=\textwidth]{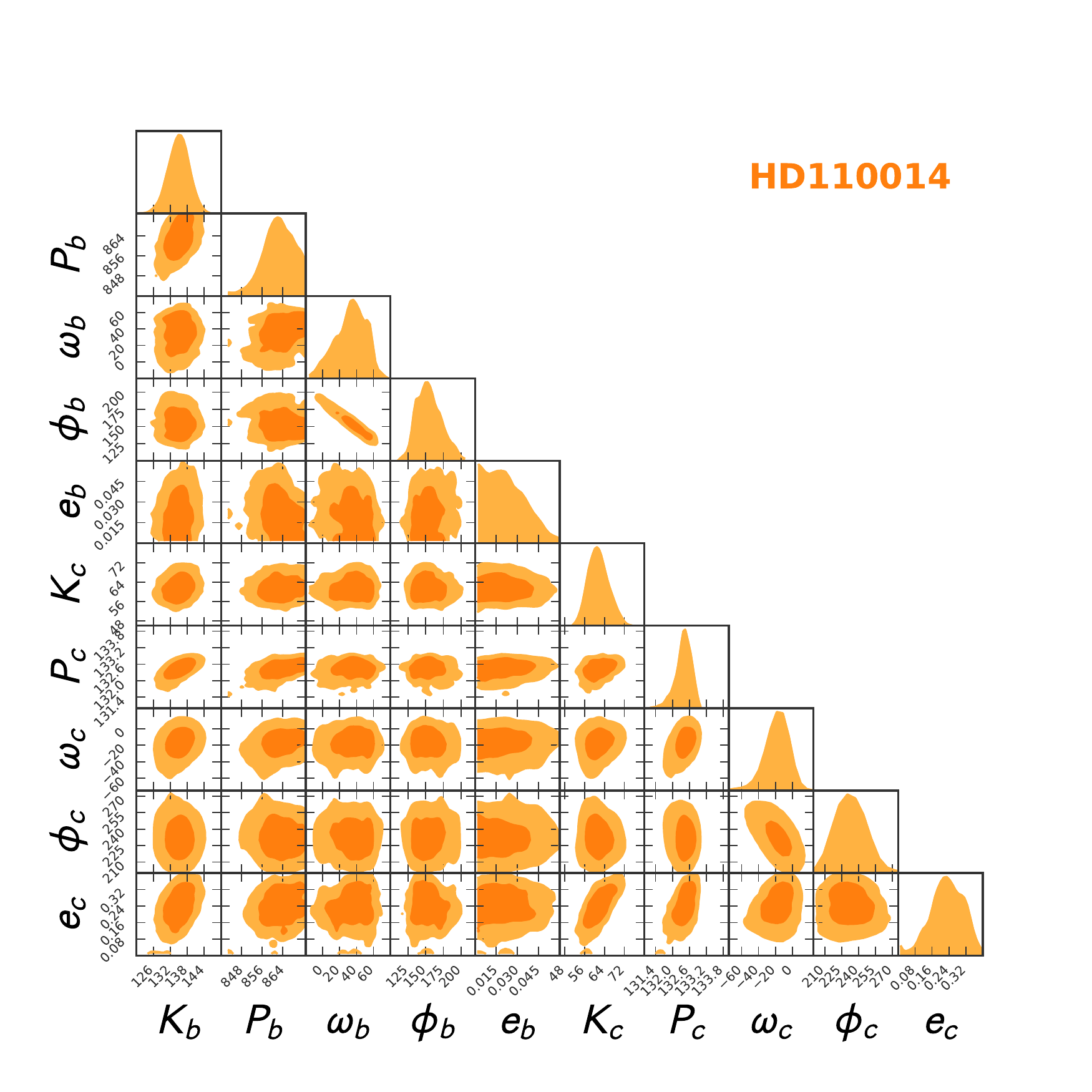}
	\caption{Bayesian posterior distributions of HD~110014 b and HD~110014 c's orbital parameters derived from {\sc astroemperor}. From left to right (top to bottom), the parameters are $K_{b}$, $P_{b}$, $\omega_{b}$, $\phi_{b}$, $e_{b}$, $K_{c}$, $P_{c}$, $\omega_{c}$, $\phi_{c}$ and $e_{c}$. Credible intervals are denoted by the solid contours with increments of 1-$\sigma$.}
    \label{fig:HD110014_confusogram}
\end{figure*}

The HD~110014 system is found to be dynamically stable, with a broad swathe of parameter space centred on the nominal solution producing system architectures that last for the full 10$^{8}$ yrs of our dynamical integrations. We show the results of the stability analysis, sampling the 3-$\sigma$ parameter space around the nominal orbital solution determined from the radial velocities in Fig. \ref{fig:HD110014_stability}. The results of the Bayesian analysis, showing what we infer to be the global best-fit parameters for the system, are presented in Fig. \ref{fig:HD110014_confusogram}.

\subsection{HD~133131A}

\begin{figure*}
    \includegraphics[width=0.45\textwidth]{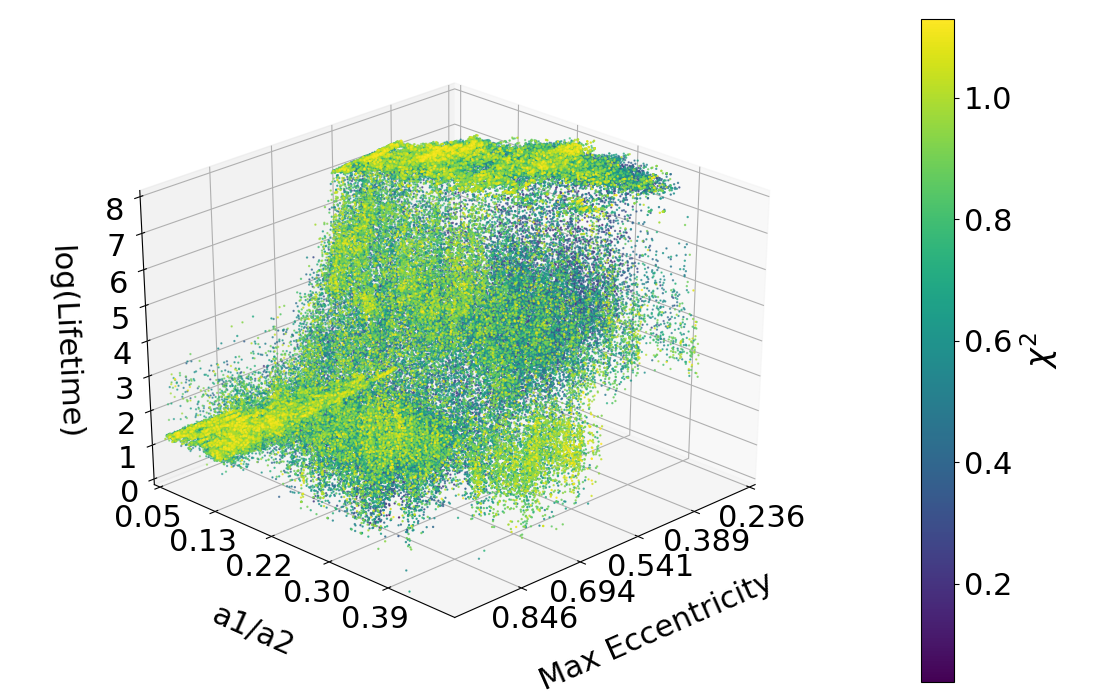}
	\includegraphics[width=0.45\textwidth]{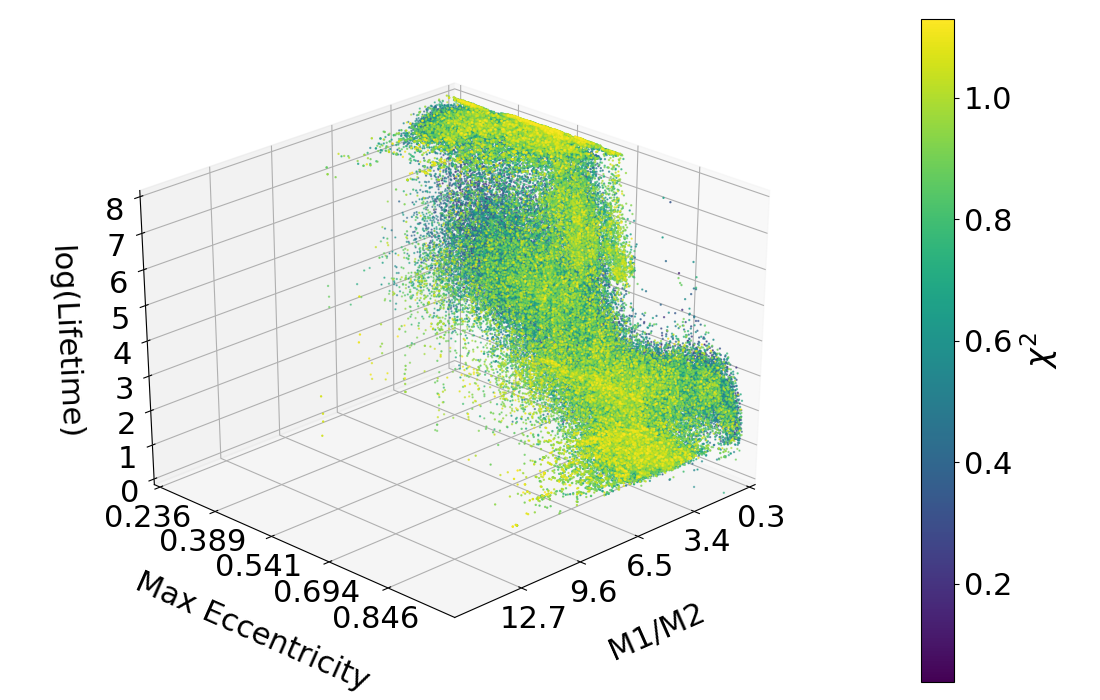}\\
	\includegraphics[width=0.45\textwidth]{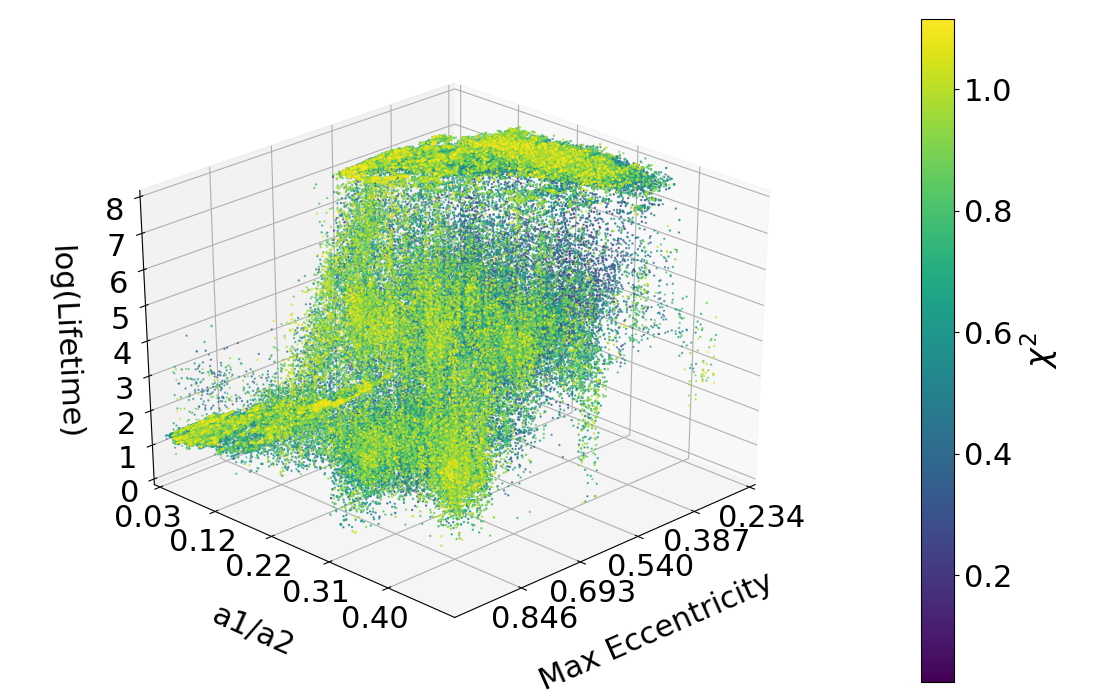}
	\includegraphics[width=0.45\textwidth]{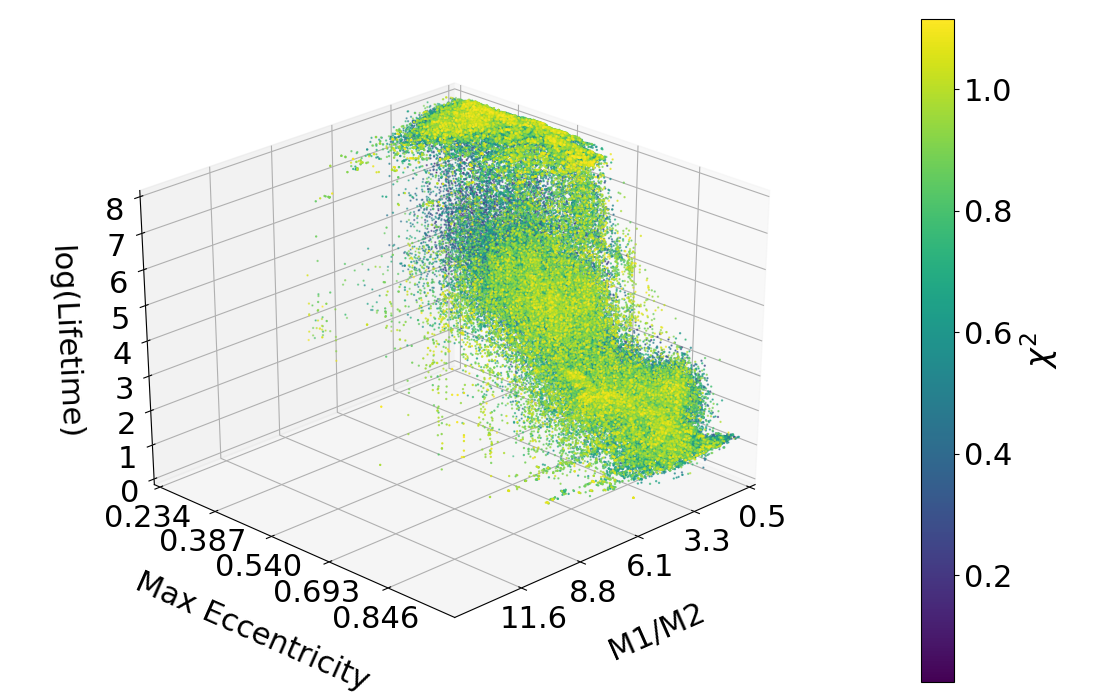}
    \caption{Plots of the dynamical stability of the HD~133131A planetary system for both the high eccentricity (top) and low eccentricity (bottom) orbital solutions. On the left we show the log(lifetime) as a function of the largest initial eccentricity fit to HD~133131Ab and c and the ratio of their orbital semi-major axes, whilst on the right we show the log(lifetime) as a function of of the largest initial eccentricity fit and the mass ratio between HD~133131Ab and c. The colour bar shows the goodness of fit ($\chi^{2}$) of each solution tested. The stability revealed by our dynamical simulations is complex, with regions of both extreme stability (log(lifetime) $\sim$ 100 Myrs) and instability (log(lifetime) $\sim$ 100 yrs) lying within the 3-$\sigma$ reach of the nominal best-fit orbital parameters.  \label{fig:HD133131A_stability}}
\end{figure*}

\begin{figure*}
	\includegraphics[width=\textwidth]{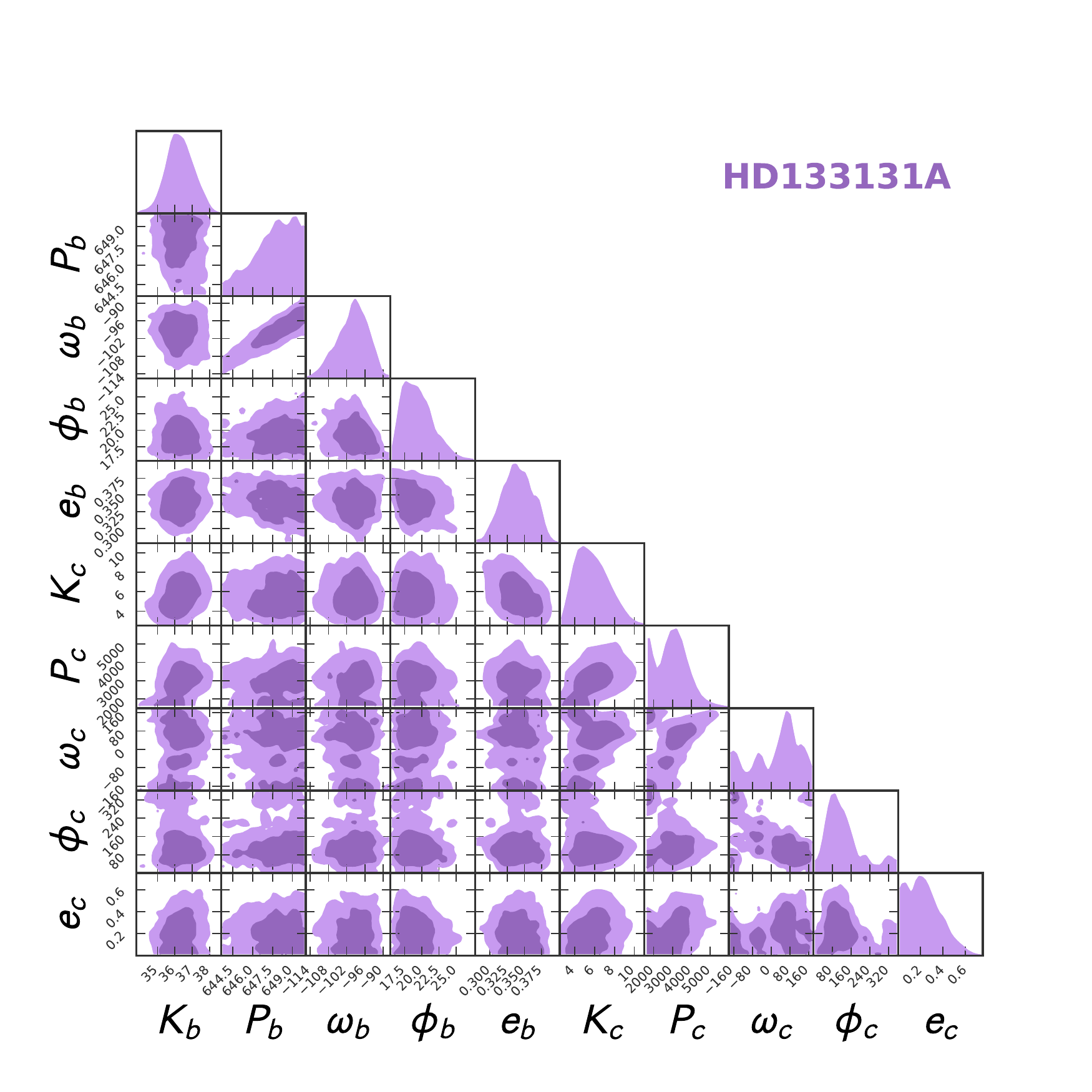}
	\caption{Bayesian posterior distributions of HD~133131A b and HD~133131A c's orbital parameters derived from {\sc astroemperor}. From left to right (top to bottom), the parameters are $K_{b}$, $P_{b}$, $\omega_{b}$, $\phi_{b}$, $e_{b}$, $K_{c}$, $P_{c}$, $\omega_{c}$, $\phi_{c}$ and $e_{c}$. Credible intervals are denoted by the solid contours with increments of 1-$\sigma$.}
    \label{fig:HD133131A_confusogram}
\end{figure*}

The HD~133131A system shows a very complex parameter space in the stability plots. As one would expect, the stability of the system generally increases towards lower orbital eccentricities and lower mass ratios between the two planetary components. The overall stability appears to be insensitive to the ratio of the semi-major axes for the planets, with long-lived solutions possible across the full range of values probed for this parameter. Interestingly, we demonstrate that stable architectures for the planetary system exist in both the high and low orbital eccentricity scenarios for the system. We show the results of the stability analysis, sampling the 3-$\sigma$ parameter space around the nominal orbital solution determined from the radial velocities in Fig. \ref{fig:HD133131A_stability}. The results of the Bayesian analysis, showing what we infer to be the global best-fit parameters for the system, are presented in Fig. \ref{fig:HD133131A_confusogram}. Further observations to refine the planet properties of this system will be required to definitively characterise its dynamical stability.

\section{Discussion}
\label{sec:Dis}

The results of our dynamical modelling for the three systems considered in this work, HD~67087, HD~110014 and HD~133131A show three distinctly different outcomes. For the first system tested, HD~67087, we find no orbital solutions that exhibit long-term dynamical stability. As a result, we are forced to conclude that, if the planets proposed to orbit that star are real, they must move on orbits significantly different from those proposed in the discovery work, and sampled in our simulations. It seems likely that new radial velocity observations of HD~67087, extending the temporal baseline over which the star has been observed, will yield fresh insights to the system -- either significantly constraining and altering the proposed orbit for the outermost planet, or even revealing that that eccentric solution is in fact the result of multiple unresolved planets at large orbital radii. Such an outcome is far from unusual -- and indeed, it is often the case that, with more data, a single eccentric planet seen in RV data is resolved to actually be two planets moving on near circular orbits \citep[e.g.][]{2013Wittenmyer,EightSingle}. For now, however, we can do no more than to call the existence of HD~67087~c into question, pending the acquisition of such additional data.

In contrast to the instability of HD~67087, our simulations of the HD~110014 system reveal that the best-fit solution for that two-planet system lies in a broad region of strong dynamical stability. In this case, our simulations simply reveal that the system, as proposed in the discovery work, is dynamically feasible -- and in a sense, the simulations add little beyond that.

The case of HD~133131A is somewhat more interesting. Here, our simulations reveal that solutions that fit the observational data can exhibit both strong dynamical stability, and extreme instability (with dynamical lifetimes of just a few years). Both the high- and low-eccentricity solutions considered in \citet{2016Teske} can produce scenarios that are stable for the full 100 Myr of our simulations. In both the high- and low-eccentricity cases, the stable solutions cluster around the least eccentric available scenarios. The more widely separated the two planets, the more eccentric their orbits can be before instability occurs -- a natural result of the stability being driven by the minimum separation between the planets, rather than their orbital semi-major axes. The more widely the semi-major axes of the orbits are spaced, the more eccentric they must be to bring the planets into close proximity. These results show once again the benefits inherent to such dynamical analysis -- reminding us how studying the dynamical evolution of a given system can help to provide stronger constraints on the orbits of the planets contained therein than is possible by studying the observational data on their own.

A comparison of our results to the analysis of the AMD stability criterion presented in \cite{2017Laskar} shows agreement between the two different techniques for the dynamical stability of the three systems. Whilst HD~67087 and HD~110014 are respectively very clear cut cases of an unstable and a stable system, HD~133131A exhibits a more complex behaviour. HD~133131A may be dynamically stable, but the inferred lifetime for the planetary system as proposed is sensitive to the chosen initial conditions; this system therefore represents an edge case of stability where limitations of available data and the respective analyses provide no clear answer to the veracity of the previously inferred planetary system.

Combining these new results with our previous dynamical analyses, as summarised in the introduction, we may consider that the AMD criterion is a reliable estimator of stability for planetary systems. There are 13 systems (out of 131 considered in that work) from \cite{2017Laskar} that have had dynamical modelling of their stability. In \cite{2017Laskar}, a planetary system is considered strongly stable if all planet-pairs have $\beta$ values less than 1, such that collisions are impossible whilst weakly stable planetary systems are those in which the inner-most planet might collide with the star without disrupting the remainder of the planetary system. In five systems, both the AMD criterion and dynamical modelling agree on their dynamical stability (HD~142, HD~159868, NN Ser (AB), GJ~832, and HD~110014); the planets in each of these systems are dynamically well separated and therefore not strongly interacting \citep[][this work]{2011Horner,2012bWittenmyer,2014bWittenmyer}. Six systems are unstable according to the AMD criterion with values of $\beta$ in the range 1 to 5 for the planet pair (HD~155358, 24 Sex, HD~200964, HD~73526, HD~33844, HD~47366), but all are in mean motion resonances and have been demonstrated to be dynamically stable through $n$-body simulations \citep{2012Robertson,2012cWittenmyer,2014aWittenmyer,2016Wittenmyer,2019Marshall}. The remaining two systems (HD~67087, HD~133131A) are dynamically unstable in both the AMD and dynamical analysis (this work). However, dynamical analysis of the HD~133131A system reveals regions of dynamical stability consistent with the observed radial velocities, prompting the need for further investigation of this system and its architecture. Neither of these two unstable planetary systems have $\beta$ values radically different from those of the planetary systems in resonance, or each other, such that determining their stability can only be carried out using dynamical simulations. The existence of such systems in the known planet population as demonstrated in our analysis therefore showcases the necessity of performing long duration dynamical analyses of proposed planetary system architectures to reveal the complex dynamical interplay between high mass planets, the evolution of their orbital elements, and determine what constraints this places on the available parameter space for the endurance of the proposed planetary system over its lifetime.

\section{Conclusions}
\label{sec:Con}

We re-analysed the dynamical stability of the exoplanet systems around HD~67087, HD~110014, and HD~133131A, using available radial velocity data. These three planetary systems have poorly constrained orbital parameters, and had previously been identified as being potentially unstable. We combine a determination of the best-fit orbital parameters from least-squares fitting to the data with $n$-body simulations to determine the global best-fit solution for the planetary system architectures, and thereafter determine the probability distribution of the orbital solutions through Bayesian inference. 

Our dynamical analysis confirms that the published planetary system parameters for HD~67087bc are dynamically unstable on very short timescales, and we must conclude that the system, as published, is dynamically unfeasible. As more data are collected for the HD~67087 system, it seems likely that the true nature of the candidate planets therein will be revealed, and that future planetary solutions for that system will veer towards dynamical stability as the planetary orbits become better constrained. 

In the case of HD~110014 bc we demonstrate that the system parameters can be dynamically stable for the full duration of our 100 Myr integrations. The third system, HD~133131A , exhibits much more complex behaviour, with HD~133131A bc being strongly unstable over much of the parameter space exhibited in this work including the region encompassing the nominal best-fit to the orbital parameters. In agreement with previous analysis of this system, we strongly disfavour a high eccentricity orbital solution for planet c. Additional observations of this system will be required to more precisely determine the planetary properties for HD~133131A bc and thereby categorically rule on the plausibility of the proposed planetary system.

These results demonstrate the complementarity of various techniques to deduce the stability of planetary systems, with good agreement between the results of our various works, and that of the AMD approach. We highlight the appropriateness of dynamical simulations for determination the long-term stability of planetary systems in the presence of strongly interacting planets, which although costly in a computing sense capture the full essence of planetary interaction in such systems which is not possible with other techniques. We finally assert that the orbital parameters for these three systems which have been determined in this work (as summarised in Table 3) should be the accepted values adopted by exoplanet archives or elsewhere. This work is thus one additional thread in the tapestry of cross-checking of published results through various means that ensures the reliability of archival information on planetary properties and the architectures of planetary systems which are essential to inform models of the formation and evolution of the exoplanet population \citep[e.g. ][]{2019Childs,2019Denham,2020He,2020VolkMalhotra}.

\section*{Acknowledgements}

This is a pre-copyedited, author-produced PDF of an article accepted for publication in MNRAS  following peer review. The version of record Marshall et al., 2020, MNRAS, 494, 2, 2280--2288 is available online \href{https://academic.oup.com/mnras/article/494/2/2280/5819459}{here}.

We thank the anonymous referee for their comments which helped to improve the article.

This research has made use of NASA's Astrophysics Data System and the SIMBAD database, operated at CDS, Strasbourg, France.

JPM acknowledges research support by the Ministry of Science and Technology of Taiwan under grants MOST104-2628-M-001-004-MY3 and MOST107-2119-M-001-031-MY3, and Academia Sinica under grant AS-IA-106-M03.

\textit{Software}: This research has made use of the following Python packages: \textsc{matplotlib} \citep{2007Hunter}; \textsc{numpy} \citep{2006Oliphant}; \textsc{pygtc} \citep{2016Bocquet}; \textsc{emcee} \citep{2013ForemanMackey}; \textsc{corner} \citep{2016ForemanMackey}; \textsc{mercury} \citep{1999Chambers}.





\bibliographystyle{mnras}
\bibliography{refs}


\bsp	
\label{lastpage}
\end{document}